\documentclass[twocolumn,showpacs,aps,prd,nofootinbib]{revtex4}
\usepackage{graphicx}
\usepackage{dcolumn}
\usepackage{amsmath}
\usepackage{epsfig}
\usepackage{colordvi}
\usepackage{color}
\usepackage{hhline}

\RequirePackage{xspace}

\newcommand{\BABARPubYear}    {13}
\newcommand{\BABARPubNumber}  {010}
\newcommand{\SLACPubNumber} {15705}
 
\usepackage{relsize}
\def\babar{\mbox{\slshape B\kern-0.1em{\smaller A}\kern-0.1em
    B\kern-0.1em{\smaller A\kern-0.2em R}}}
     
\mathchardef\Upsilon="7107
\def\Y#1S{\ensuremath{\Upsilon{(#1S)}}\xspace}

\def\pep2{PEP-II}

\long\def\inst#1{\par\nobreak\kern 4pt\nobreak
  {\it #1}\par\vskip 10pt plus 3pt minus 3pt}
  
\begin{document}

\begin{flushleft}
SLAC-PUB-\SLACPubNumber \\
\babar-PUB-\BABARPubYear/\BABARPubNumber \\
\end{flushleft}

\title{\large \bf
\boldmath
Measurement of the $e^+e^-\to p\bar{p}$ cross
section in the energy range from 3.0 to 6.5 GeV
}

%
\author{J.~P.~Lees}
\author{V.~Poireau}
\author{V.~Tisserand}
\affiliation{Laboratoire d'Annecy-le-Vieux de Physique des Particules (LAPP), Universit\'e de Savoie, CNRS/IN2P3,  F-74941 Annecy-Le-Vieux, France}
\author{E.~Grauges}
\affiliation{Universitat de Barcelona, Facultat de Fisica, Departament ECM, E-08028 Barcelona, Spain }
\author{A.~Palano$^{ab}$ }
\affiliation{INFN Sezione di Bari$^{a}$; Dipartimento di Fisica, Universit\`a di Bari$^{b}$, I-70126 Bari, Italy }
\author{G.~Eigen}
\author{B.~Stugu}
\affiliation{University of Bergen, Institute of Physics, N-5007 Bergen, Norway }
\author{D.~N.~Brown}
\author{L.~T.~Kerth}
\author{Yu.~G.~Kolomensky}
\author{M.~J.~Lee}
\author{G.~Lynch}
\affiliation{Lawrence Berkeley National Laboratory and University of California, Berkeley, California 94720, USA }
% ...
\collaboration{The \babar\ Collaboration}
\noaffiliation

\begin{abstract}
The $e^+e^-\to p\bar{p}$ cross section and the proton magnetic
form factor have been measured in the center-of-mass 
energy range from 3.0 to 6.5 GeV using the initial-state-radiation 
technique with an undetected photon.
This is the first measurement of the form factor at energies higher
than 4.5 GeV. The analysis is based on 469~fb$^{-1}$ of
integrated luminosity collected with the \babar\ detector 
at the PEP-II collider at $e^+e^-$ center-of-mass energies near 10.6~GeV.
The branching fractions for the decays $J/\psi \to p\bar{p}$ and 
$\psi(2S) \to p\bar{p}$ have also been measured. 
\end{abstract}

\pacs{13.66.Bc, 14.20.Dh, 13.40.Gp, 13.25.Gv}

\maketitle

\setcounter{footnote}{0}

\section{ \boldmath Introduction\label{intro}}
\begin{figure}
\includegraphics[width=.4\textwidth]{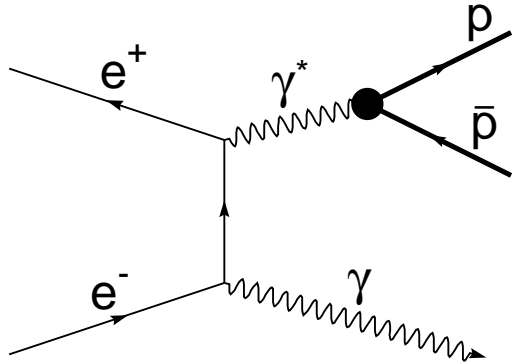}
\caption{The Feynman diagram for the ISR process $e^+e^-\to p\bar{p}\gamma$.}
\label{fig1}
\end{figure}

In this paper we analyze the initial-state-radiation (ISR)
process $e^+e^-\to p\bar{p}\gamma$ represented by Fig.~\ref{fig1}. 
This analysis is a continuation of our previous 
studies~\cite{ppbabar,ppbabarn}, where the ISR technique was used to 
measure the cross section of the nonradiative process $e^+e^-\to p\bar{p}$ 
over the center-of-mass (c.m.) energy range from $p\bar{p}$  threshold, 
$2m_p c^2=1.88$ GeV, up to 4.5 GeV. In Refs.~\cite{ppbabar,ppbabarn} it is 
required that the ISR photon be detected (large-angle ISR).
In this paper, we analyze events
in which the ISR photon is emitted along the $e^+e^-$ collision axis
(small-angle ISR) and is therefore not detected.
This allows us to increase the detection
efficiency for ISR events with $p\bar{p}$ invariant mass above
3.2 GeV/$c^2$, to select $p\bar{p}\gamma$
events with lower background, and, therefore, to extend the 
energy range for measurement of the $e^+e^-\to p\bar{p}$ cross section.
A discussion of the difference between the large- and small-angle ISR 
techniques is given in Ref.~\cite{ISRrev}.

The Born cross section for the ISR process integrated over the nucleon momenta
and the photon polar angle is given by
\begin{equation}
\frac{{\rm d}\sigma_{e^+e^-\to p\bar{p}\gamma}(M_{p\bar{p}})}
{{\rm d}M_{p\bar{p}}} = 
\frac{2M_{p\bar{p}}}{s}\, W(s,x)\,\sigma_{p\bar{p}}(M_{p\bar{p}}),
\label{eq1}
\end{equation}
where $M_{p\bar{p}}$ is the $p\bar{p}$ invariant mass, $s$ is the $e^+e^-$ 
c.m.~energy squared,
$x\equiv{E_{\gamma}^\ast}/\sqrt{s}=1-{M_{p\bar{p}}^2}/{s}$, 
and $E_{\gamma}^\ast$ is the ISR photon energy in the $e^+e^-$ 
c.m.~frame\footnote{Throughout this paper,
the asterisk denotes quantities in the $e^+e^-$ c.m.~frame;
all other variables are given in the
laboratory frame.}.
The function~\cite{ISRrev}
\begin{equation}
W(s,x) = \frac{\alpha}{\pi x}(\ln{\frac{s}{m_e^2}}-1)(2-2x+x^2)
\label{eq2}
\end{equation}
specifies the probability of ISR photon emission, where $\alpha$ is the 
fine structure constant and $m_e$ is the electron mass. 
Equations (\ref{eq1}) and (\ref{eq2}) describe ISR processes at lowest QED 
order. To calculate the function $W(x)$ more precisely, taking into account
higher-order diagrams involving loops and extra photon emission, we make use 
of the analytic techniques described in Refs.~\cite{radf1,radf2,radf3} and 
the Monte Carlo (MC) generator of ISR events, Phokhara~\cite{phokhara}.

The cross section for $e^+e^-\to p\bar{p}$ is given by
\begin{equation}
\sigma_{p\bar{p}}(M_{p\bar{p}}) = \frac{4\pi\alpha^{2}\beta C}{3M_{p\bar{p}}^2}
\left [|G_M(M_{p\bar{p}})|^{2} + \frac{2m_p^2}{M_{p\bar{p}}^2}
|G_E(M_{p\bar{p}})|^{2}\right],
\label{eq4}
\end{equation}
where $\beta =\sqrt{1-4m_p^2/M_{p\bar{p}}^2}$, 
$C = y/(1-e^{-y})$ is the Coulomb 
correction factor~\cite{Coulomb}, and $y = {\pi\alpha}(1+\beta^2)/\beta$.
The Coulomb factor makes the cross section nonzero at threshold.
The cross section depends on the magnetic ($G_M$) and 
electric ($G_E$) form factors. 
At large $p\bar{p}$ invariant masses the second term in Eq.~(\ref{eq4}) is 
suppressed as ${2m_p^2}/{M_{p\bar{p}}^2}$, and therefore the measured total cross section
is not very sensitive
to the value of the electric form factor.  The value of the magnetic form
factor can be extracted from the measured cross section with relatively
small model uncertainty using, for example, 
the assumption that $|G_M|=|G_E|$~\cite{BES,CLEO,NU}.

The existing experimental data on $|G_M(M_{p\bar{p}})|$ at high $p\bar{p}$ invariant 
masses were obtained in $e^+e^-$~\cite{ppbabarn,BES,CLEO,NU} and $p\bar{p}$
annihilation~\cite{E760,E835}. At energies higher than 3 GeV the 
value of the magnetic form factor decreases rapidly with increasing energy.
The energy dependence measured in Refs.~\cite{ppbabarn,CLEO,E760,E835} agrees
with the dependence $\alpha_s^2(M_{p\bar{p}}^2)/M_{p\bar{p}}^4$
predicted by QCD for the asymptotic proton form factor~\cite{QCD}. 
However,  the two precision measurements of Ref.~\cite{NU} based
on CLEO data indicate that the decrease of the form factor at energies
near 4~GeV is somewhat slower.

In this work we improve the accuracy of our measurements of the
$e^+e^-\to p\bar{p}$ cross section and of the proton magnetic form factor for 
$p\bar{p}$ invariant masses greater than 3 GeV/$c^2$, and extend the range 
of measurement up to 6.5 GeV/$c^2$.

\section{ \boldmath The \babar\ detector, data and simulated samples}
\label{detector}
We analyse a data sample corresponding to an integrated luminosity of 
469~fb$^{-1}$~\cite{babar-lum}
recorded with the  \babar\ detector~\cite{ref:babar-nim} at the SLAC \pep2\ 
asymmetric-energy (9-GeV $e^-$ and 3.1-GeV $e^+$) collider.
About 90\% of the data were collected at an $e^+e^-$ c.m.~energy of 10.58~GeV 
(the $\Upsilon$(4S) mass), and the remainder at 10.54 GeV. 

Charged-particle tracking is
provided by a five-layer silicon vertex tracker (SVT) and
a 40-layer drift chamber (DCH), operating in the 1.5 T 
magnetic field of a superconducting solenoid. The transverse momentum resolution
is 0.47\% at 1~GeV/$c$. The position and energy of a photon-produced
cluster are measured with a CsI(Tl) electromagnetic calorimeter. 
Charged-particle identification (PID) is provided by specific ionization 
measurements in the SVT and DCH, and by an internally reflecting 
ring-imaging Cherenkov detector. Muons are identified in
the solenoid's instrumented flux return.

The events of the process under study and the background processes
$e^+e^-\to \pi^+\pi^-\gamma$, $K^+K^-\gamma$, and $\mu^+\mu^-\gamma$
are simulated with the Phokhara~\cite{phokhara} event generator,
which takes into account next-to-leading-order radiative corrections.
To estimate the model uncertainty of our measurement, 
the simulation for the signal process is performed under two form-factor 
assumptions, namely $|G_M|=|G_E|$ and $|G_E|=0$. 
To obtain realistic estimates of pion and kaon backgrounds,
the experimental values of the pion and kaon electromagnetic form
factors measured by the CLEO Collaboration at $\sqrt{s}=3.67$ 
GeV~\cite{CLEO} are used in the event generator. The invariant-mass 
dependence of the 
form factors is assumed to be $1/m^2$, according to the QCD 
prediction for the asymptotic behavior of the form factors~\cite{QCDpion}.
The $e^+e^-\to e^+e^-\gamma$ process is simulated with
the BHWIDE~\cite{BHWIDE} event generator.

Background from the two-photon process $e^+e^-\to e^+e^-p\bar{p}$ is
simulated with the GamGam event generator~\cite{gamgam}.
In addition, possible background contributions from $e^+e^-\to q\bar{q}$, where 
$q$ represents a $u$, $d$ or $s$ quark, are simulated with the JETSET~\cite{JETSET} event
generator. Since JETSET also generates ISR events, it can be used to
study background from ISR processes with extra $\pi^0$'s, such as
$e^+e^-\to p\bar{p}\pi^0\gamma$, $p\bar{p}\pi^0\pi^0\gamma$, etc.
The most important non-ISR background process, $e^+e^-\to p\bar{p}\pi^0$,
is simulated separately~\cite{ppbabar}.

The detector response is simulated using the Geant4~\cite{GEANT4} package.
The simulation takes into account the variations in the detector and
beam-background conditions over the running period of the experiment.

\section{ \boldmath Event selection}\label{sel}
We select events with two charged-particle tracks with opposite 
charge originating from the interaction region. Each track must 
have transverse momentum greater than 0.1~GeV/$c$, be in the polar angle
range $25.8^\circ <\theta < 137.5^\circ$, and be identified 
as a proton or antiproton. The pair of proton and antiproton candidates is 
fit to a common vertex with a beam-spot constraint, and the $\chi^2$ 
probability for this fit is required to exceed 0.1\%.
\begin{figure}
\includegraphics[width=.4\textwidth]{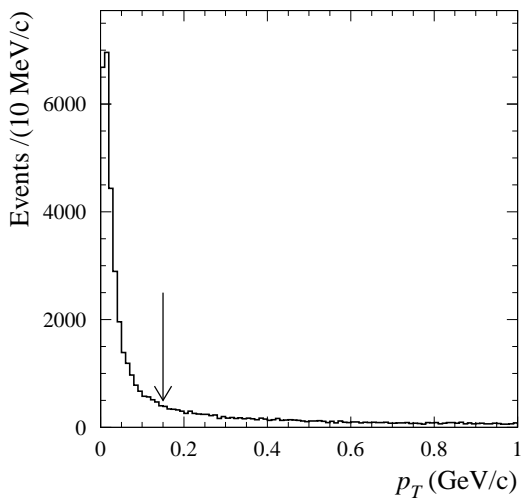}
\caption{The distribution of the $p\bar{p}$ transverse
momentum for simulated $e^+e^-\to p\bar{p}\gamma$ events.
The arrow indicates $p_{\rm{T}}=0.150$ GeV/$c$.
\label{fig2}}
\end{figure}
\begin{figure}
\includegraphics[width=.4\textwidth]{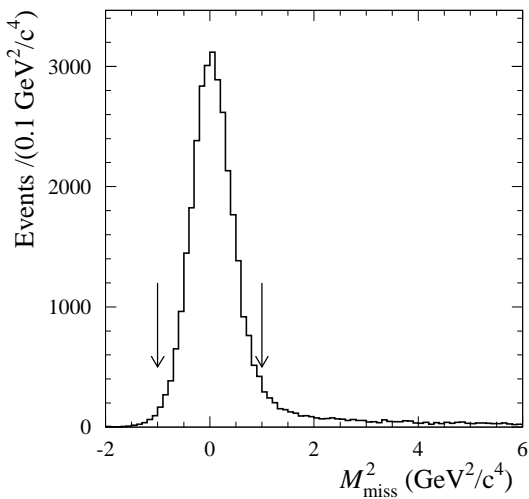}
\caption{The $M^2_{\rm{miss}}$ distribution
for simulated $e^+e^-\to p\bar{p}\gamma$ events, where
$M^2_{\rm{miss}}$ is the missing-mass-squared recoiling against the
$p\bar{p}$ system. The arrows indicate 
$|M_{\rm{miss}}^2|=1$ GeV$^2/c^4$.
\label{fig3}}
\end{figure}

The final event selection is based on two variables: the $p\bar{p}$ transverse
momentum ($p_{\rm{T}}$) and the missing-mass-squared ($M^2_{\rm{miss}}$)
recoiling against the $p\bar{p}$ system. The $p_{\rm{T}}$ distribution
for simulated $e^+e^-\to p\bar{p}\gamma$ events is shown in Fig.~\ref{fig2}.
The peak near zero corresponds to ISR photons emitted along
the collision axis, while the long tail is due to photons emitted at large angles.
We apply the condition $p_{\rm{T}}<0.15$ GeV/$c$, which removes large-angle
ISR events, and strongly suppresses background from the process
$e^+e^-\to p\bar{p}\pi^0$ and from ISR processes with extra $\pi^0$'s.
The process $e^+e^-\to p\bar{p}\pi^0$ was the dominant background source at
large invariant masses in our previous studies of the $e^+e^-\to p\bar{p}\gamma$ process
with large-angle ISR~\cite{ppbabar,ppbabarn}.

In the $e^+e^-$ c.m.~frame protons with low $p\bar{p}$ invariant masses
are produced in a narrow cone around
the vector opposite to the ISR photon direction. Due to limited detector
acceptance, the low-mass region cannot be studied with small-angle ISR.
A $p\bar{p}$ pair with $p_{\rm{T}}<0.15$ GeV/$c$ is detected in \babar\
when its invariant mass is larger than 3.0 (4.5) GeV/$c^2$ for
an ISR photon emitted along the electron (positron) beam direction.
The corresponding average proton or antiproton  momentum in the laboratory 
frame is about 2 (5) GeV/$c$.
The difference between the two photon directions arises from
the energy asymmetry of the $e^+e^-$ collisions at PEP-II. 
Since particle misidentification probability strongly increases
at large momentum, we reject events with the ISR photon emitted along the
positron beam. This condition decreases the detection efficiency by about 20\%
for signal events with invariant masses above 5 GeV/$c^2$.

The missing-mass-squared distribution for simulated $e^+e^-\to p\bar{p}\gamma$
events is shown in Fig.~\ref{fig3}. We select events with
$|M_{\rm{miss}}^2|<1$ GeV$^2/c^4$. This condition suppresses background
from two-photon and ISR events, which have large positive $M_{\rm{miss}}^2$, and
background from $e^+e^-\to e^+e^-\gamma$, $\mu^+\mu^-\gamma$
events, which have negative $M_{\rm{miss}}^2$.
Sideband regions in $M_{\rm{miss}}^2$ and in $p_{\rm{T}}$ for
ISR background are used to estimate remaining background contributions
from these sources.

\begin{figure}
\includegraphics[width=.4\textwidth]{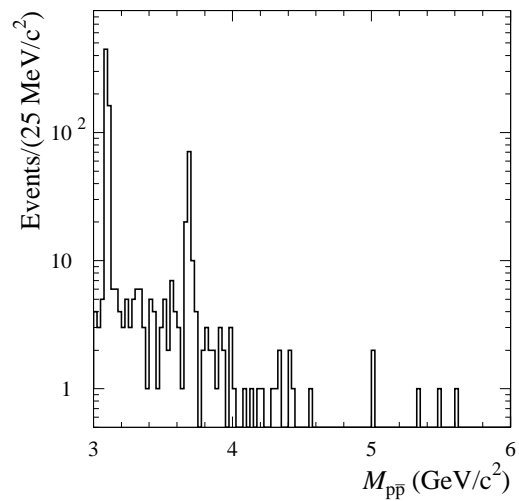}
\caption{The $p\bar{p}$ invariant-mass spectrum for
selected data $p\bar{p}\gamma$ candidates.
\label{fig4}}
\end{figure}
The $p\bar{p}$ invariant-mass spectrum for the selected data candidates
is shown in Fig.~\ref{fig4}. The total number of selected events
is 845. About 80\% of selected events originate from
$J/\psi\to p\bar{p}$ and $\psi(2S)\to p\bar{p}$ decays.
We do not observe events with invariant mass above 6 GeV/$c^2$.

\section{Background estimation and subtraction}\label{background}
 The processes $e^+e^-\to \pi^+\pi^-\gamma$, $K^+K^-\gamma$,
$\mu^+\mu^-\gamma$, and $e^+e^-\gamma$ in which the charged particles are
misidentified as protons, are potential sources of background in the sample of 
selected data events. In addition, the two-photon process 
$e^+e^-\to e^+e^-p\bar{p}$, and processes with protons and neutral particles
in the final state, such as
$e^+e^-\to p \bar{p} \pi^0$ and $e^+e^-\to p \bar{p} \pi^0 \gamma$, may
yield background contributions.

\subsection{\boldmath Background from 
$e^+e^-\to \pi^+\pi^-\gamma$, $e^+e^-\to K^+ K^-\gamma$, 
$e^+e^-\to \mu^+\mu^-\gamma$, and $e^+e^-\to e^+e^-\gamma$}
In Ref.~\cite{ppbabarn} it was shown that the \babar\ MC simulation reproduces
reasonably well the probability for a pion or a kaon to be identified as
a proton. Consequently, the simulation is used to estimate the
$e^+e^-\to \pi^+\pi^-\gamma$  and $e^+e^-\to K^+K^-\gamma$ background
contributions in the present analysis. No events satisfying the selection criteria 
for $p\bar{p}\gamma$ are observed in the $\pi^+\pi^-\gamma$ and $K^+K^-\gamma$ MC samples.
Since these MC samples exceed those expected for pion and kaon events
in data by about an order of magnitude,
we conclude that these background sources can be neglected.

To estimate possible electron and muon background a method based on the 
difference in the $M_{\rm{miss}}^2$ distributions for signal and background
events is used. For $e^+e^-\to \mu^+\mu^-\gamma$ events the ratio of the 
number of events with $|M_{\rm{miss}}^2|<1$ GeV$^2/c^4$ to the number 
with $M_{\rm{miss}}^2<-1$ GeV$^2/c^4$ varies from 0.03 to about 0.1 in 
the $M_{p\bar{p}}$ range of interest. Smaller values are expected for 
$e^+e^-\to e^+e^-\gamma$ events.
In data we observe 15 events with $M_{\rm{miss}}^2<-1$ GeV$^2/c^4$, 
of which 6 events are expected to originate from signal (of these, 5 are from 
$J/\psi\to p\bar{p}$ and $\psi(2S)\to p\bar{p}$ decays).
From the ratio values given above, we estimate that muon and electron
background in our selected event sample does not exceed 1 event. 
The estimated background contributions for different invariant-mass intervals
are listed in Table~\ref{tab2}.
\begin{table}
\caption{The number of selected $p\bar{p}\gamma$ candidates ($N_{\rm data}$) and
the estimated numbers of background events  from the processes
$e^+e^-\to \mu^+\mu^-\gamma$ and 
$e^+e^-\to e^+e^-\gamma$ ($N_{\ell\ell\gamma}$),
$e^+e^-\to e^+e^- p\bar{p}$ ($N_{2\gamma}$),
and the ISR processes with extra neutral particle(s), such as
$e^+e^-\to p\bar{p}\pi^0\gamma$, $p\bar{p}2\pi^0\gamma$
($N_{\rm bkg}^{\rm ISR}$). In the invariant-mass intervals 
3.0--3.2 GeV/$c^2$ and 3.6--3.8 GeV/$c^2$
the contribution of the $J/\psi\to p\bar{p}$ and $\psi(2S)\to p\bar{p}$
decays are subtracted (see Sec.~\ref{psi}), with related statistical 
uncertainties reported.
\label{tab2}}
\begin{ruledtabular}
\begin{tabular}{ccccc}
$M_{p\bar{p}}$ (GeV/$c^2$) & $N_{\rm data}$ & $N_{\ell\ell\gamma}$ & 
$N_{2\gamma}$ & $N_{\rm bkg}^{\rm ISR}$ \\
\hline
\\[-2.1ex]
3.0--3.2 & $35\pm7$ & $<0.1$ & $0.5\pm0.4$ & $1.5\pm0.6$ \\
3.2--3.4 & 32  & $<0.1$ & $0.5\pm0.3$   & $1.3\pm0.6$ \\
3.4--3.6 & 31  & $<0.1$ & $0.15\pm0.10$ & $0.7\pm0.5$ \\
3.6--3.8 & $17\pm5$ & $<0.1$ & $0.20\pm0.10$ & $0.0\pm0.2$ \\
3.8--4.0 & 16  & $<0.1$ & $0.10\pm0.04$ & $0.7\pm0.4$ \\
4.0--4.5 & 12  & $<0.1$ & $0.10\pm0.03$ & $0.7\pm0.4$ \\
4.5--5.5 & 5   & $<0.3$ & $0.05\pm0.02$ & $1.0\pm0.5$ \\
5.5--6.5 & 1   & $<0.3$ & $ < 0.01    $ & $0.4\pm0.3$ \\
\end{tabular}
\end{ruledtabular}
\end{table}

\subsection{Two-photon background}
\begin{figure}
\includegraphics[width=.4\textwidth]{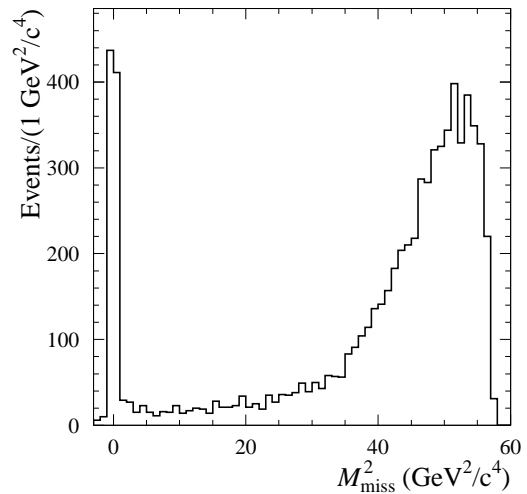}
\caption{The $M_{\rm{miss}}^2$ distribution
for data events selected using all the criteria described in Sec.~\ref{sel}
except $|M_{\rm{miss}}^2|<1$ GeV$^2/c^4$.
\label{fig5}}
\end{figure}
Figure~\ref{fig5} shows the $M_{\rm{miss}}^2$ distribution for data
events selected using all the criteria described in Sec.~\ref{sel}
except $|M_{\rm{miss}}^2|<1$ GeV$^2/c^4$. Events with large recoil mass
arise from the two-photon process
$e^+e^-\to e^+e^-\gamma^\ast\gamma^\ast\to e^+e^-p\bar{p}$.
The two-photon background in the region $|M_{\rm{miss}}^2|<1$ GeV$^2/c^4$
is estimated from the number of data events with
$M_{\rm{miss}}^2>d$ using the scale factor
$R_{\gamma\gamma}=N_{\gamma\gamma}(|M_{\rm{miss}}^2|<1)/N_{\gamma\gamma}(M_{\rm{miss}}^2>d)$
obtained from the $e^+e^-\to e^+e^- p\bar{p}$ simulation. Since the $M_{\rm{miss}}^2$ distribution
for two-photon events changes with $p\bar{p}$ invariant mass, the parameter 
$d$ is changed from 40 GeV$^2/c^4$ for the invariant-mass interval 
3.0--3.2 GeV/$c^2$ to 15 GeV$^2/c^4$ for the interval 5.5--6.5 GeV/$c^2$. 
To determine a realistic value of the scale factor, the simulated events are reweighted according
to the proton angular distribution observed in data. 
The value of the scale factor is found to increase from $5\times 10^{-4}$ in the
3.0--3.2 GeV/$c^2$ interval to $2\times 10^{-2}$ in the 5.5--6.5 GeV/$c^2$ interval. 
Fortunately, the number of observed two-photon events decreases significantly
over this same range. The estimated number of two-photon background events for
each invariant-mass interval is listed in Table~\ref{tab2}. The background 
is found to be small, at the level of 1\%.

\subsection{ISR background}
\begin{figure*}
\includegraphics[width=.32\textwidth]{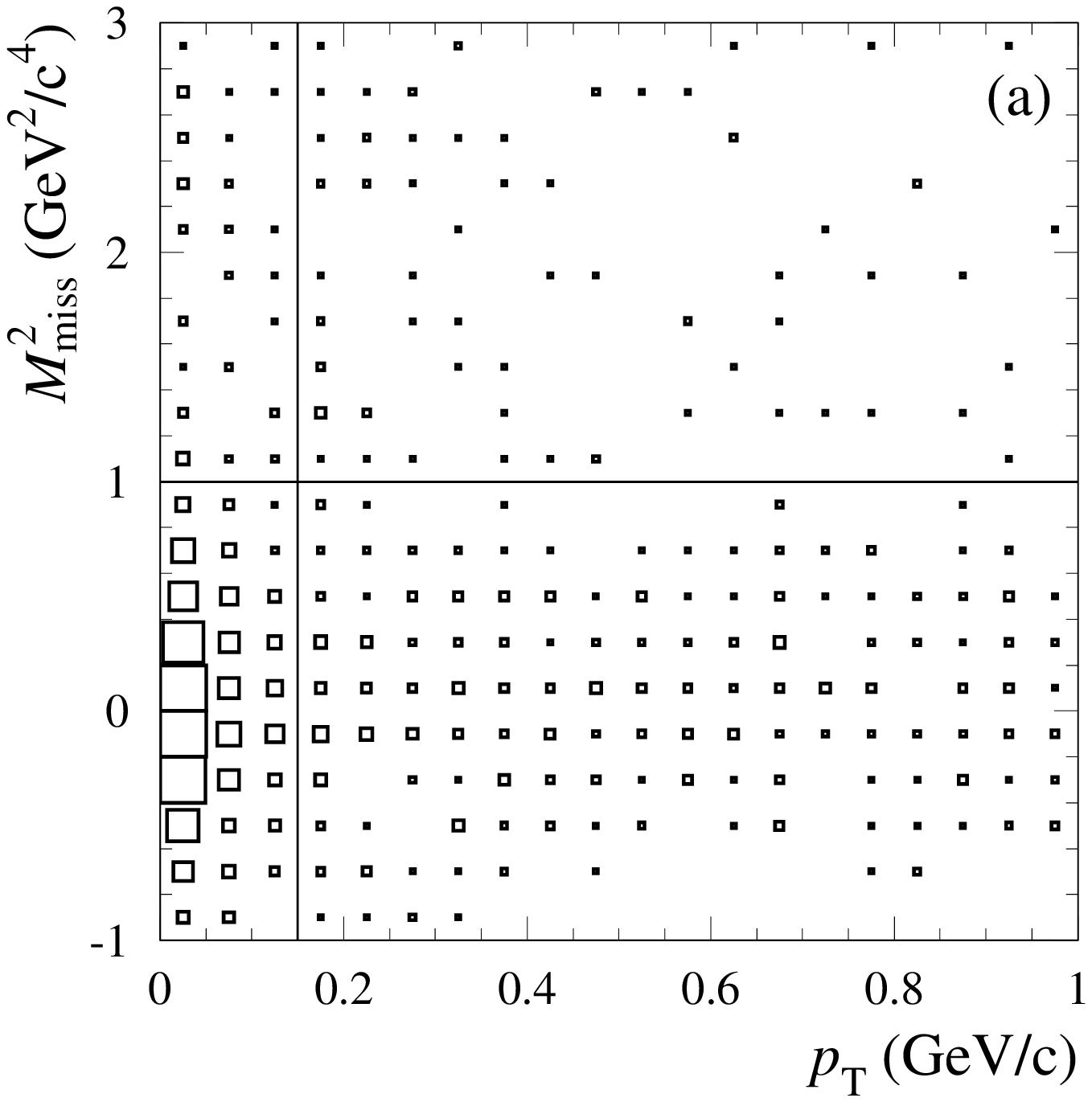}
\includegraphics[width=.32\textwidth]{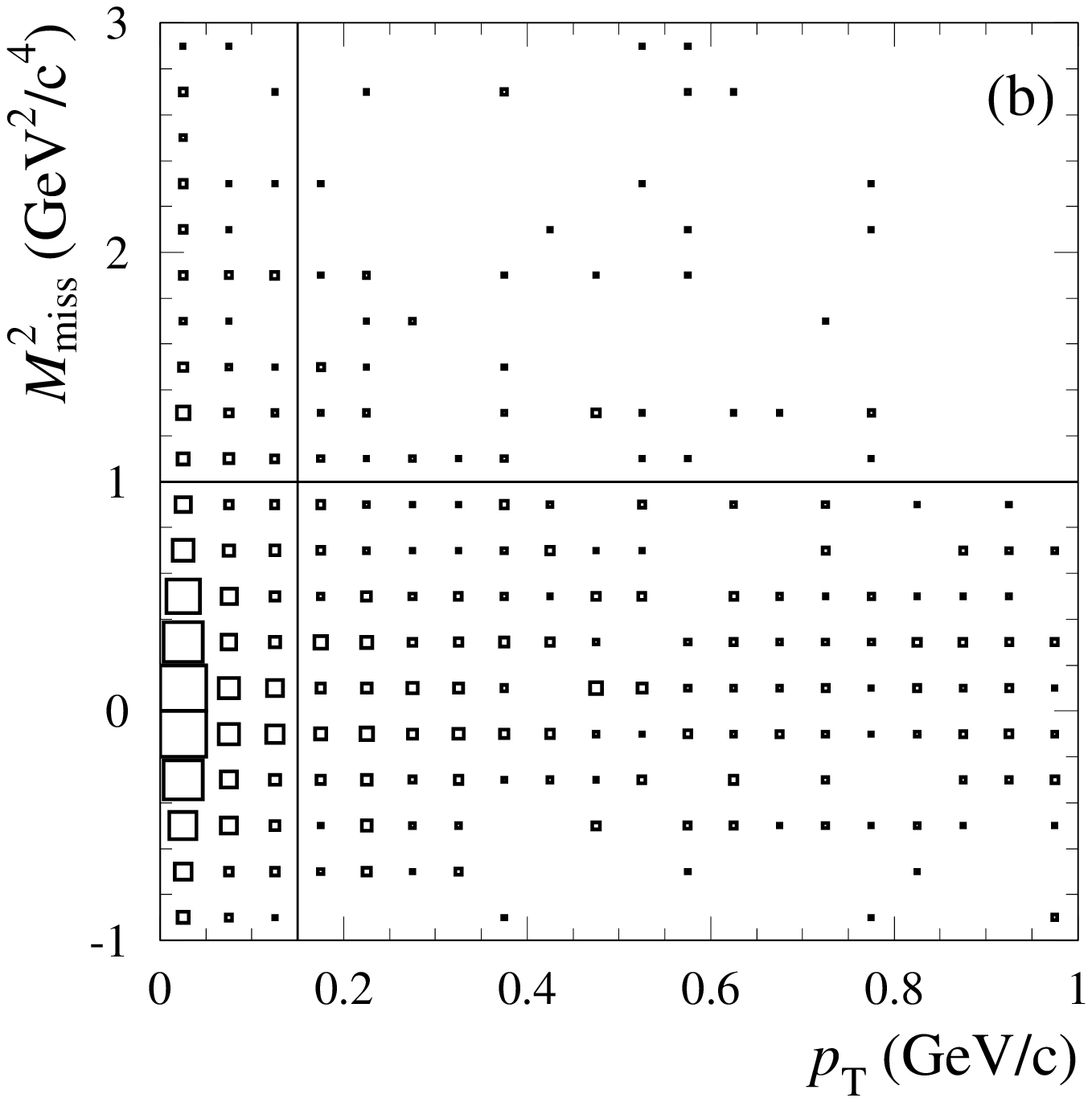}
\includegraphics[width=.32\textwidth]{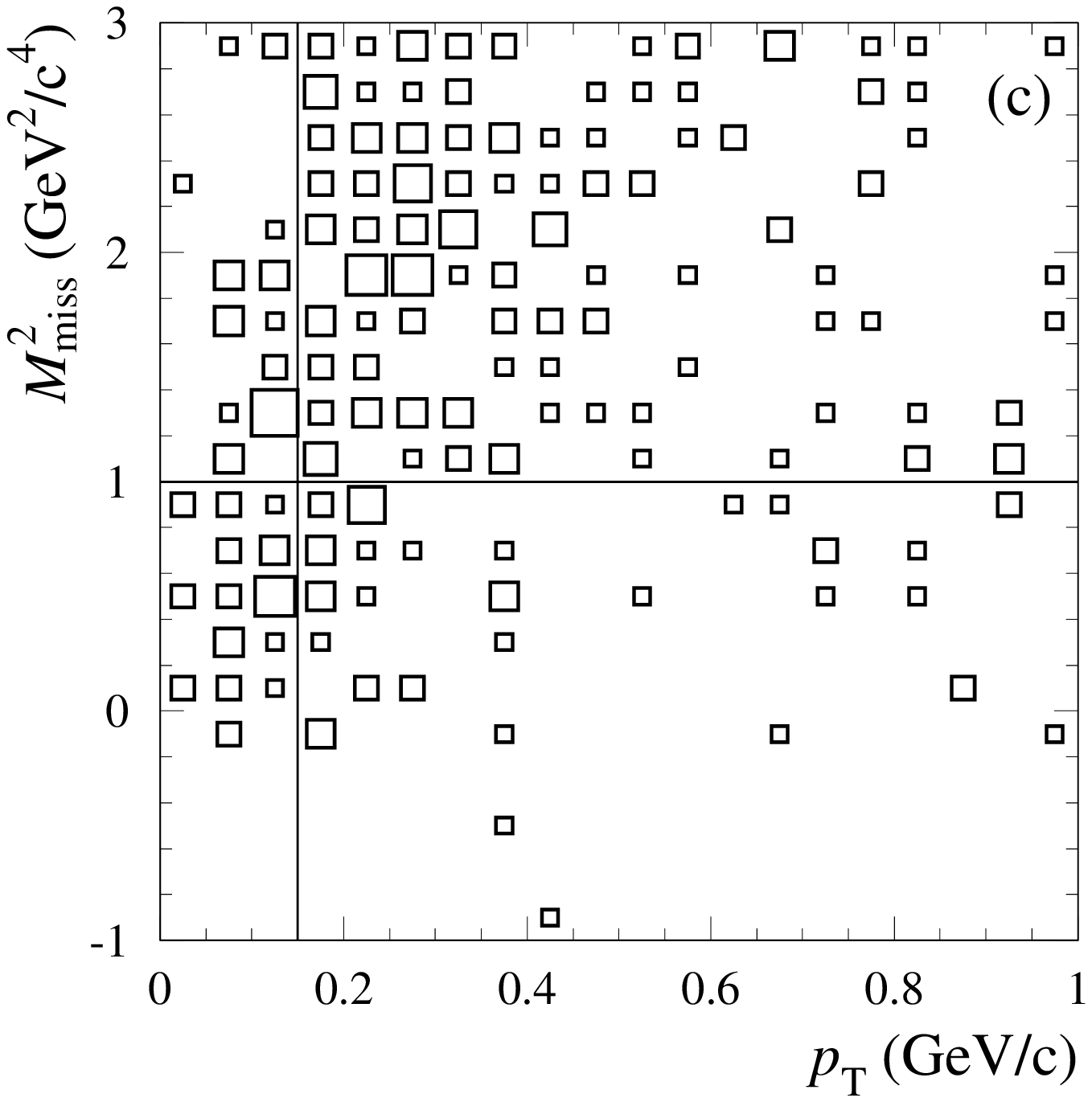}
\caption{The distributions of $M_{\rm{miss}}^2$ versus $p_T$
(a) for data events with $M_{p\bar{p}} > 3.2$ GeV/$c^2$, (b) for simulated
signal events, and (c) for simulated ISR background events.
\label{fig6}}
\end{figure*}
To estimate background from ISR processes with at least one extra neutral
particle, such as $e^+e^-\to p\bar{p}\pi^0\gamma$,
$e^+e^-\to p\bar{p}\eta\gamma$, $e^+e^-\to p\bar{p}\pi^0\pi^0\gamma$, etc.,
we use differences in the $p_T$ and $M_{\rm{miss}}^2$ distributions
for signal and background events. Figure~\ref{fig6} shows
the two-dimensional distributions of $M_{\rm{miss}}^2$ versus $p_T$
for data events with $M_{p\bar{p}} > 3.2$ GeV/$c^2$, and for simulated
signal and ISR background events. The ISR background is simulated
using the JETSET event generator. It should be noted that most of the background
events (about 90\%) shown in Fig.~\ref{fig6} arise from $e^+e^-\to
p\bar{p}\pi^0\gamma$. The lines in Fig.~\ref{fig6} indicate the boundaries
of the signal region (bottom left rectangle) and of the sideband 
region (top right rectangle).
The number of data events in the sideband ($N_2$) is used to estimate
the number of background events in the signal region by using
\begin{equation}
N_{\rm bkg}^{\rm ISR}=\frac{N_2-\beta_{sig}N_1}{\beta_{bkg}-\beta_{sig}},
\end{equation}
where $N_1$ is the number of data events in the signal region, and
$\beta_{sig}$ and $\beta_{bkg}$ are the $N_2/N_1$ ratios for the signal
and background, respectively. These ratios are determined from
MC simulation to be $\beta_{sig}=0.043\pm0.002$ and $\beta_{bkg}=5\pm 1$.
Both coefficients are found to be practically independent of
$p\bar{p}$ invariant mass. The estimated numbers of ISR background events for 
different invariant-mass regions are listed in Table~\ref{tab2}. This is the 
main source of background for the process under study.

The background from the process $e^+e^-\to p\bar{p}\pi^0$, which
was the dominant background source in our previous large-angle
studies~\cite{ppbabar,ppbabarn}, is found to be negligible 
in the data sample selected with the criteria for small-angle ISR events.

\section{Detection efficiency\label{deteff}}
\begin{figure}
\includegraphics[width=.4\textwidth]{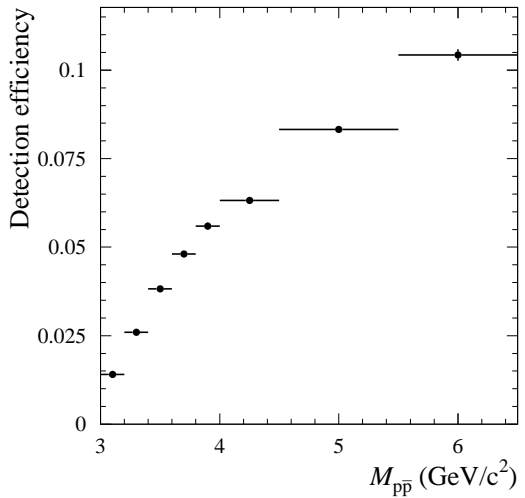}
\caption{The $p\bar{p}$ invariant-mass dependence of the detection
efficiency obtained from MC simulation in the model with $|G_E|=|G_M|$.
\label{fig7}}
\end{figure}

\begin{figure}
\includegraphics[width=.4\textwidth]{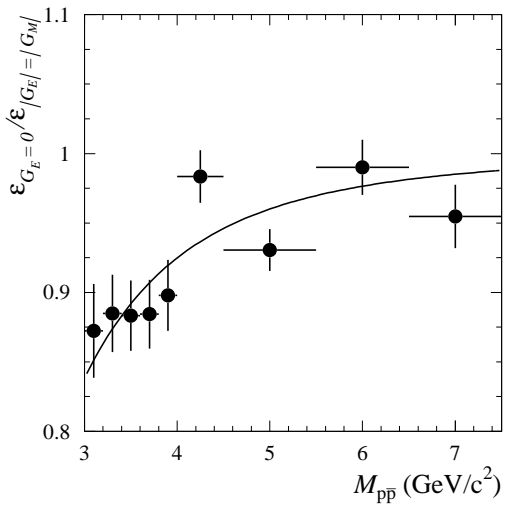}
\caption{The ratio of the detection efficiencies obtained from MC simulation
using $G_E=0$ and using $|G_E|=|G_M|$. The solid curve is drawn to guide 
the eye.
\label{fig8}}
\end{figure}

The detection efficiency determined using MC simulation is
shown in Fig.~\ref{fig7} as a function of $p\bar{p}$ invariant mass.
The efficiency is calculated under the assumption that $|G_E|=|G_M|$.
To study the model dependence of the detection efficiency, we
analyze a sample of MC events
produced using a model with $G_E=0$. The ratio of the efficiencies
obtained in the two models is shown in Fig~\ref{fig8}. The
deviation of this ratio from unity is taken as an estimate
of the model uncertainty on the detection efficiency.

The efficiency determined from MC simulation
($\varepsilon_{MC}$) must be corrected to account for
data-MC simulation differences in detector response according to
\begin{equation}
\varepsilon=\varepsilon_{MC}\prod (1+\delta_i),
\label{eq_eff_cor}
\end{equation}
where the $\delta_i$ are the efficiency corrections listed 
in Table~\ref{tab_ef_cor}.
\begin{table}
\caption{The values of the efficiency corrections, $\delta_i$.
\label{tab_ef_cor}}
\begin{ruledtabular}
\begin{tabular}{lc}
Source               &$\delta_i$ (\%) \\
\hline
\\[-2.1ex]
Track reconstruction & $ 0.0\pm0.5$  \\
Nuclear interaction  & $ 1.1\pm0.4$ \\
PID                  & $-1.9\pm2.0$ \\
Conditions on $p_T$ and $M^2_{\rm miss}$   & $4.3\pm2.6$ \\
\hline
\\[-2.1ex]
Total                & $3.5\pm3.3$ \\
\end{tabular}
\end{ruledtabular}
\end{table}
The corrections for data-MC simulation differences in track reconstruction,
nuclear interaction, and PID were estimated in our previous
publications~\cite{ppbabar,ppbabarn}. 
Systematic effects on $p_T$ and $M^2_{\rm miss}$ may bias 
the estimated efficiency through the selection criteria.
This is studied using
$e^+e^-\to J/\psi\gamma \to p\bar{p}\gamma$ events. In Sec.~\ref{psi} 
the number of $J/\psi$ events is determined with the requirements
$p_T < 1$ GeV/$c$ and $-2 < M^2_{\rm miss} <3$ GeV$^2/c^4$, which are
significantly looser than our standard criteria.
The double data-MC simulation ratio of the numbers of $J/\psi$ events selected
with the standard and looser criteria, $1.043 \pm 0.026$, is used
to estimate the efficiency correction. The corrected values of the detection
efficiency are listed in Table~\ref{sumtab}.

\section{\boldmath $J/\psi$ and $\psi(2S)$ decays into $p\bar{p}$}\label{psi}
\begin{figure}
\includegraphics[width=.4\textwidth]{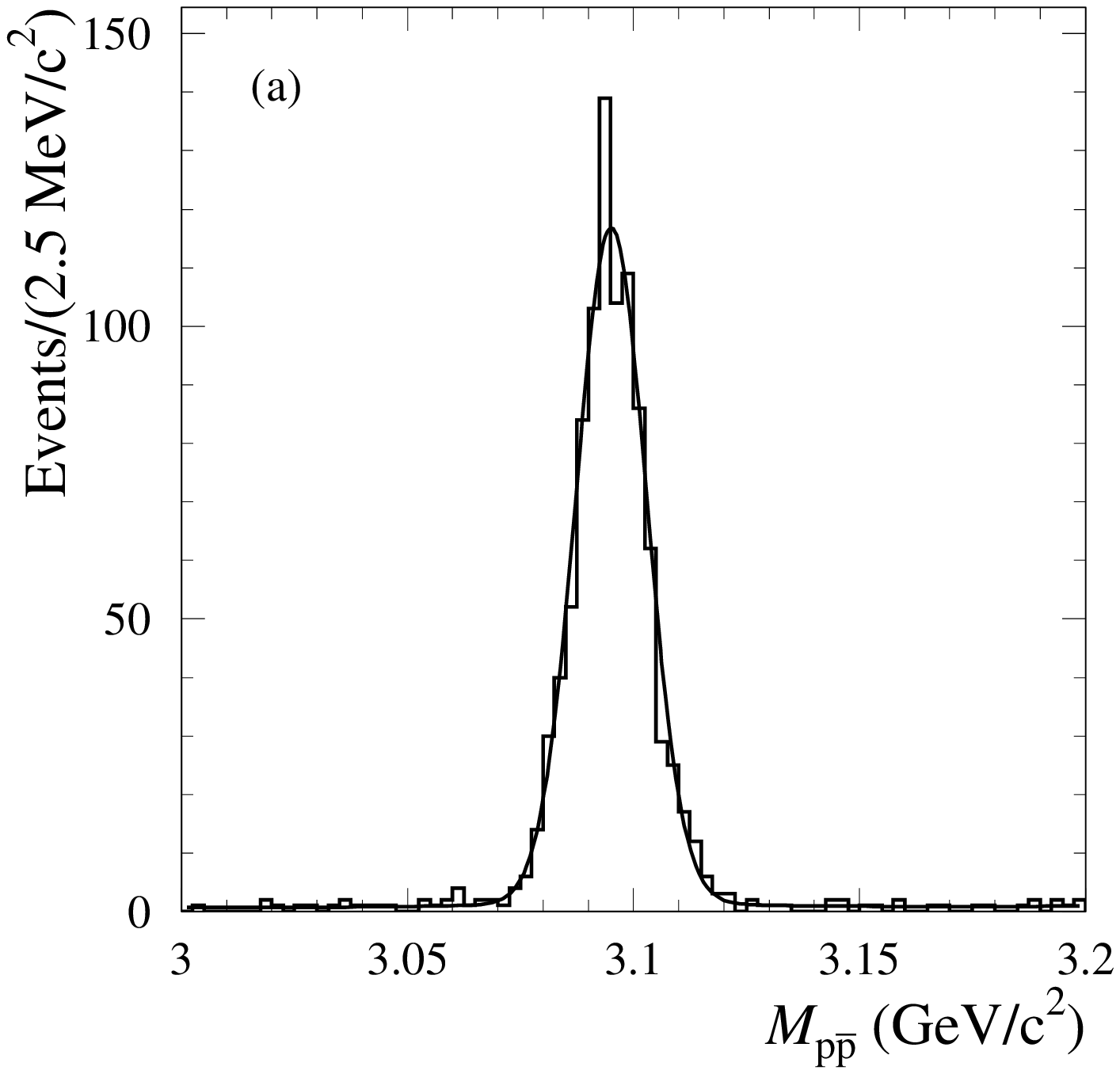}
\includegraphics[width=.4\textwidth]{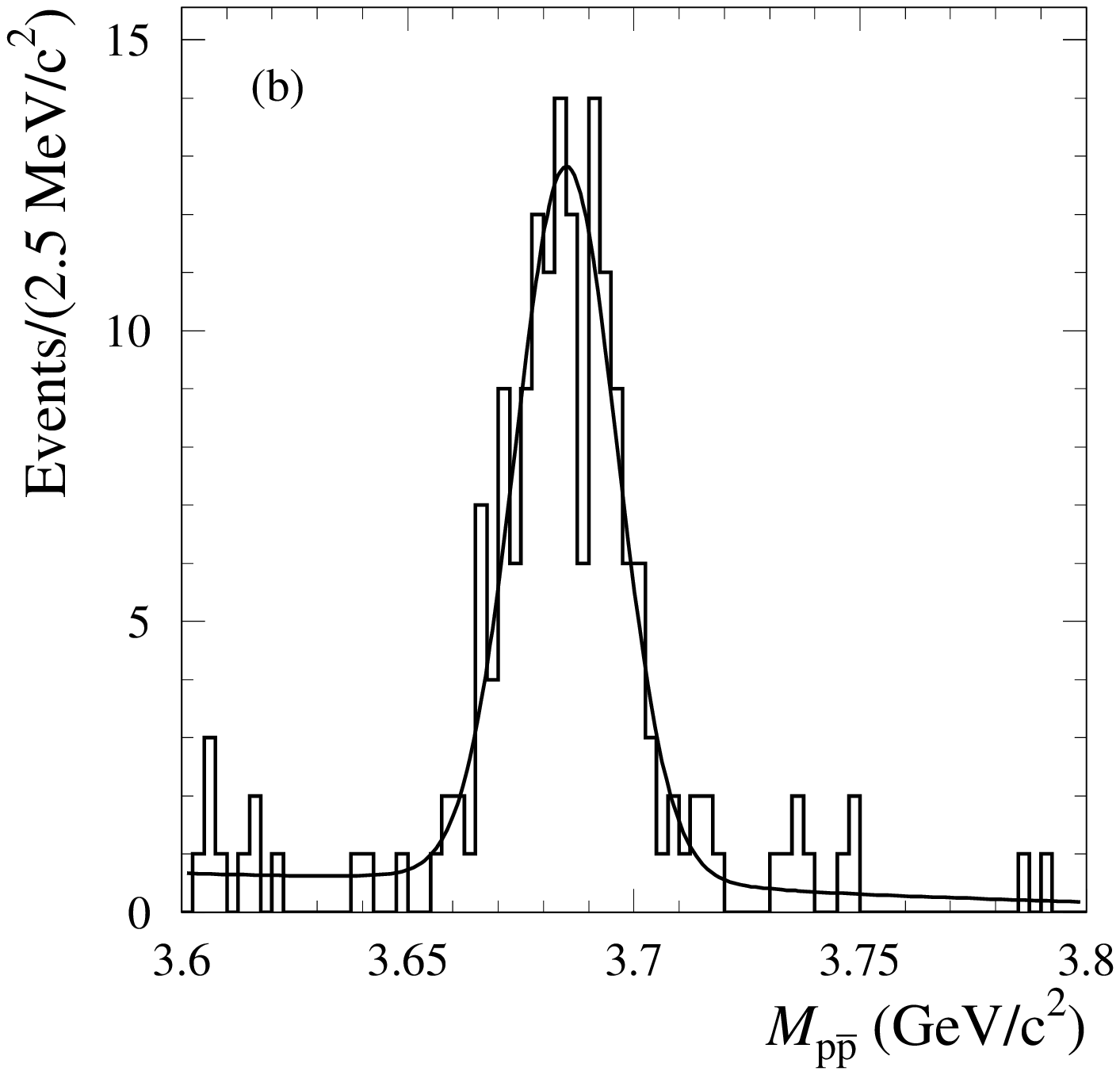}
\caption{The $p\bar{p}$ invariant-mass spectrum in the invariant-mass region
near (a) the $J/\psi$, and (b) the $\psi(2S)$. The curves show the
results of the fits described in the text.}
\label{fig9}
\end{figure}
The $p\bar{p}$ invariant-mass spectra for selected events in the $J/\psi$ and
$\psi(2S)$ invariant-mass regions are shown in Fig.~\ref{fig9}.
The events are selected by requiring
$p_T < 1$ GeV/$c$ and $-2 < M^2_{\rm miss} <3$ GeV$^2/c^4$.
To determine the number of resonance events, both spectra
are fitted using the sum of a probability density function (PDF)
for resonance events and a linear background function. The resonance PDF is
a Breit-Wigner function convolved with
a double-Gaussian function describing detector resolution.
The parameters of the resolution function
are determined from simulation. To account for possible differences
in detector response between data and simulation,
the simulated resolution function is modified by allowing
an additional $\sigma_G$ to be added in quadrature to both
$\sigma$'s of the double-Gaussian function and by introducing
the possibility of an invariant-mass shift. The free parameters in the
fit to the $J/\psi$ invariant-mass region are the number of resonance events,
the total number of nonresonant background events, the slope
of the background function, $\sigma_G$, and the mass shift parameter.
In the $\psi(2S)$ fit, $\sigma_G$ is fixed to
the value obtained from the $J/\psi$ fit.
The result of the fit for the $J/\psi$ region is shown by
the solid curve in Fig.~\ref{fig9}(a), and the corresponding signal yield is
$918\pm31$ events. Similarly, the solid curve in Fig.~\ref{fig9}(b) shows the
fit result for the $\psi(2S)$ region, with signal yield of $142\pm13$ events.

The detection efficiency is estimated from MC simulation.
The event generator uses the experimental data on the polar-angle
distribution of the proton in $\psi \to p\bar{p}$ decay.
The distribution is described by the function $1+a \cos^2\vartheta$
with $a=0.595\pm0.019$ for $J/\psi$~\cite{BESjpsi}
and $0.72\pm 0.13$ for $\psi(2S)$~\cite{E835psi,BESpsi2s}. The model
error on the detection efficiency due to the uncertainty of
$a$ is estimated to be 1.5\% for the $J/\psi$ and 5\% for
the $\psi(2S)$. The efficiencies ($\varepsilon_{MC}$) are found to be
$(2.20 \pm 0.02)\%$ for the $J/\psi$ and
$(6.86 \pm 0.04)\%$ for the $\psi(2S)$.
The data-MC simulation differences discussed earlier are used to correct
the above efficiency values by $(-0.8\pm 2.1)\%$ (Table~\ref{tab_ef_cor}, 
corrections 1--3).

The value of the cross section for the production of the $J/\psi$ or
$\psi(2S)$ followed by its decay to $p\bar{p}$ is given by $N/(\varepsilon L)$,
where $N$ is the number of signal events extracted in the fit shown in 
Fig.~\ref{fig9}(a) or
Fig.~\ref{fig9}(b), $\varepsilon$ is the relevant detection efficiency, and
$L$ is the nominal integrated luminosity. The cross section values obtained in
this way are $(89.5\pm3.0\pm2.8)$ fb and  $(4.45\pm0.41\pm0.25)$ fb for the $J/\psi$
and $\psi(2S)$, respectively, where the first error is statistical and the 
second systematic.

These values correspond to the integral of the right-hand
side of Eq.~(\ref{eq1}) over the resonance lineshape, {\it i.e.} for
resonance $R$
\begin{equation}
\sigma_{\rm meas}=\int \frac{2m}{s} W(s,x) \sigma_R(m) {\rm d}m,
\end{equation}
where $m$ runs over the resonance region. For a narrow resonance
\begin{equation}
\sigma_{\rm meas}=W(s,x_R) \frac{12\pi^2}{s}
\frac{\Gamma(R\to e^+e^-)\,{\cal B}(R\to p\bar{p})}{m_R}
\end{equation}
is a very good approximation, where $x_R=1-m^2_R/s$, and $m_R$ is
the resonance mass.

From the measured values of the cross section 
we thus obtain:
\begin{eqnarray}
{\Gamma(J/\psi\to e^+e^-)\,{\cal B}(J/\psi\to p\bar{p})=} \nonumber \\
(12.9\pm 0.4\pm 0.4)\mbox{ eV}, \nonumber \\
{\Gamma(\psi(2S)\to e^+e^-)\,{\cal B}(\psi(2S)\to p\bar{p})=}\nonumber \\
(0.74\pm0.07\pm0.04)\mbox{ eV}.
\end{eqnarray}
The systematic error includes the uncertainties of the detection efficiency,
the integrated luminosity (1\%), and the theoretical uncertainty on the
production cross section (1\%).

Using the nominal values of the $e^+e^-$ widths~\cite{pdg},
the $\psi\to p\bar{p}$ branching fractions are calculated to be
\begin{eqnarray}
{\cal B}(J/\psi\to p\bar{p})=(2.33\pm0.08\pm0.09)\times 10^{-3},\nonumber \\
{\cal B}(\psi(2S)\to p\bar{p})=(3.14\pm0.28\pm0.18)\times 10^{-4}.
\end{eqnarray}
These values are in agreement with the corresponding nominal
values~\cite{pdg}, $(2.17\pm 0.07)\times 10^{-3}$ and
$(2.76\pm 0.12)\times 10^{-4}$, and with the recent BESIII
measurement~\cite{BESjpsi}
${\cal B}(J/\psi\to p\bar{p})=(2.11\pm 0.03)\times 10^{-3}$.

\section{\boldmath The $e^+e^-\to p\bar{p}$
cross section and the proton form factor\label{xsec}}
\begin{table*}
\caption{ The $p\bar{p}$ invariant-mass interval ($M_{p\bar{p}}$),
number of selected events ($N$) after background subtraction,
detection efficiency ($\varepsilon$),
ISR luminosity ($L$), measured  $e^+e^-\to p\bar{p}$ cross section
($\sigma_{p\bar{p}}$), and the proton magnetic form factor
($|G_M|$). The quoted uncertainties are statistical.
The systematic uncertainty is 4\% for the cross section, and
2\% for the form factor. The model uncertainty for the cross
section (form factor) is 15 (8)\% at 3 GeV, decreases to 5 (3)\% at 4.5 GeV,
and does not exceed 5(3)\% at higher values.
\label{sumtab}}
\begin{ruledtabular}
\begin{tabular}{cccccc}
$M_{p\bar{p}}$ (GeV/$c^2$)&$N$&$\varepsilon$ (\%)&$L$ (pb$^{-1}$)&$\sigma_{p\bar{p}}$ (pb)&$|G_M|$\\[0.3ex]
\hline
\\[-2.1ex]
3.0--3.2& $33.0\pm 7.0$ &  1.45&  271& $8.4\pm 1.8$      &$0.0310^{+0.0031}_{-0.0035}$\\ \\[-2.1ex]
3.2--3.4& $30.0\pm 5.7$ &  2.69&  292& $3.8\pm 0.7$      &$0.0221^{+0.0020}_{-0.0022}$\\ \\[-2.1ex]
3.4--3.6& $30.0\pm 5.6$ &  3.95&  314& $2.42\pm 0.45$    &$0.0186^{+0.0017}_{-0.0018}$\\ \\[-2.1ex]
3.6--3.8& $16.4\pm 5.1$ &  4.97&  337& $0.98\pm 0.30$    &$0.0124^{+0.0018}_{-0.0021}$\\ \\[-2.1ex]
3.8--4.0& $15.0\pm 4.0$ &  5.79&  361& $0.72\pm 0.19$    &$0.0112^{+0.0014}_{-0.0016}$\\ \\[-2.1ex]
4.0--4.5& $11.0\pm 3.5$ &  6.54& 1018& $0.165\pm 0.053$  &$0.0058^{+0.0009}_{-0.0010}$\\ \\[-2.1ex]
4.5--5.5& $ 4.0\pm 2.3$ &  8.62& 2637& $0.018\pm 0.010$  &$0.0022^{+0.0006}_{-0.0008}$\\ \\[-2.1ex]
5.5--6.5& $ 0.6\pm 1.1$ & 10.79& 4079& $0.0014\pm 0.0025$&$0.0007^{+0.0005}_{-0.0007}$\\
\end{tabular}
\end{ruledtabular}
\end{table*}
\begin{figure}
\includegraphics[width=.4\textwidth]{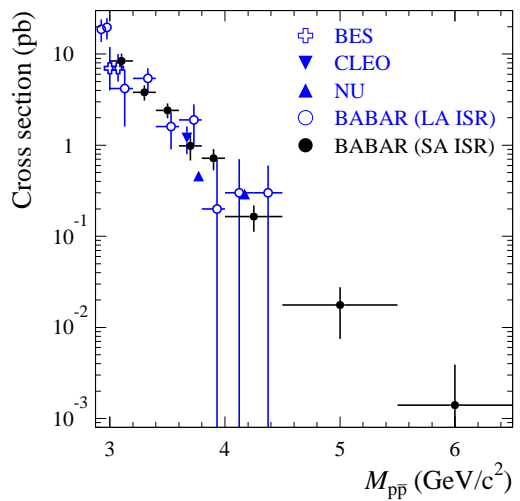}
\caption{The $e^+e^-\to p\bar{p}$ cross section
measured in this analysis [BABAR (SA ISR)] and in
other experiments: BES~\cite{BES}, CLEO~\cite{CLEO}, NU~\cite{NU},
and BABAR (LA ISR)~\cite{ppbabarn}.
\label{fig10}}
\end{figure}
The cross section for $e^+e^-\to p\bar{p}$ in each
$p\bar{p}$ invariant-mass interval $i$ is calculated as 
$N_i/(\varepsilon_i L_i)$.
The number of selected events ($N_i$) for each $p\bar{p}$ invariant-mass 
interval after background subtraction is listed in Table~\ref{sumtab}.
The values of the $L_i$ (Table~\ref{sumtab}) have been obtained by integration
of $W(s,x)$ from Refs.~\cite{radf1,radf2} over each invariant-mass interval.
They can be calculated also using the
Phokhara event generator~\cite{phokhara}. The results of
the two calculations agree within 0.5\%, which
coincides with the estimated theoretical accuracy of
the Phokhara generator~~\cite{phokhara}.
The obtained values of the $e^+e^-\to p\bar{p}$ cross section are listed
in Table~\ref{sumtab}. For the invariant-mass intervals
3.0--3.2 GeV/$c^2$ and 3.6--3.8 GeV/$c^2$ we quote the nonresonant
cross sections with the respective $J/\psi$ and $\psi(2S)$ contributions excluded.
The quoted errors are statistical, as obtained from the uncertainty in
the number of selected $p\bar{p}\gamma$ events. The systematic uncertainty is 
independent of invariant mass and is equal to 4\%. It includes the statistical
error of the detection efficiency (2\%), the uncertainty of the efficiency
correction (3.3\%), the uncertainty in the integrated luminosity (1\%), and
an uncertainty in the ISR luminosity (0.5\%). The model uncertainty
due to the unknown $|G_E/G_M|$ ratio (see Fig.~\ref{fig8}) is about
15\% at 3 GeV/$c^2$, decreases to 5\% at 4.5 GeV/$c^2$,
and does not exceed 5\% at higher values.
The measured $e^+e^-\to p\bar{p}$ cross section is shown in
Fig.~\ref{fig10} together with the results of previous $e^+e^-$
measurements.
\begin{figure}
\includegraphics[width=.48\textwidth]{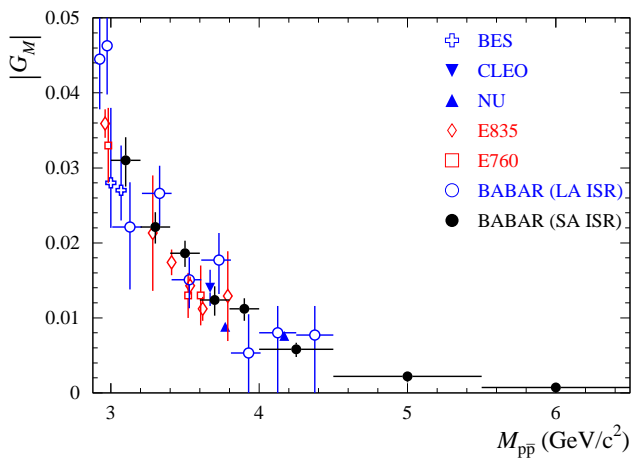}
\caption{The proton magnetic form factor measured
in this analysis [BABAR (SA ISR)] and in
other experiments: BES~\cite{BES}, CLEO~\cite{CLEO}, NU~\cite{NU},
E835~\cite{E835}, E760~\cite{E760}, BABAR (LA ISR)~\cite{ppbabarn}
\label{fig11}}
\end{figure}
\begin{figure}
\includegraphics[width=.4\textwidth]{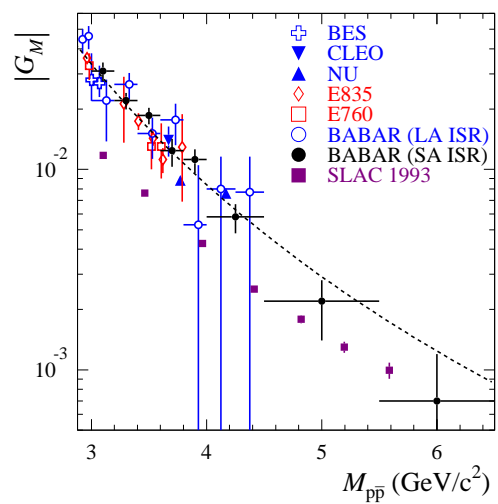}
\caption{The proton magnetic form factor measured in this
analysis [BABAR (SA ISR)] and in
other experiments: BES~\cite{BES}, CLEO~\cite{CLEO}, NU~\cite{NU},
E835~\cite{E835}, E760~\cite{E760}, BABAR (LA ISR)~\cite{ppbabarn}. Points
denoted by ``SLAC 1993'' represent data on the space-like magnetic
form factor obtained in $ep$ scattering~\cite{SLAC93} as a
function of $\sqrt{-q^2}$, where $q^2$ is the momentum transfer
squared. The curve is the result of the QCD-motivated fit
described in the text.
\label{fig12}}
\end{figure}

The values of the proton magnetic form factor are obtained using
Eq.~(\ref{eq4}) under the assumption that $|G_E|=|G_M|$. They are listed in
Table~\ref{sumtab} and shown in Fig.~\ref{fig11} (linear scale) and in 
Fig.~\ref{fig12} (logarithmic scale). It is seen that our results are in 
good agreement with the results from other experiments.
The curve in Fig.~\ref{fig12} is the result of a fit
of the asymptotic QCD dependence of the proton form factor~\cite{QCD},
$|G_M|\sim\alpha_s^2(M_{p\bar{p}}^2)/M_{p\bar{p}}^4 
\sim D/(M_{p\bar{p}}^4\log^2(M_{p\bar{p}}^2/\Lambda^2))$, to all the
existing data with $M_{p\bar{p}} > 3$ GeV/$c^2$, excluding the two points
from Ref.~\cite{NU}. Here $\Lambda=0.3$ GeV and $D$ is
a free fit parameter. The data are well described by this
function, with $\chi^2/\nu=17/24$, where $\nu$ is the number of degrees of
freedom. Including the points from Ref.~\cite{NU} in the fit increases 
$\chi^2/\nu$ to $54/26$. 

In Fig.~\ref{fig12} we also show the space-like $|G_M|$ data 
(``SLAC 1993'' points)
obtained in Ref.~\cite{SLAC93}. The QCD prediction is that the space- and time-like
asymptotic values be the same. In the region from 3.0 to 4.5 GeV/$c^2$ 
the value of the time-like form factor is about two times larger 
than that of the space-like one.
Our points above 4.5 GeV/$c^2$ give some indication that the difference
between time- and space-like form factors may be decreasing, although
our measurement uncertainties are large in this region.

\section{Summary} \label{summary}
The process $e^+e^-\to p\bar{p}\gamma$ has been studied in
the $p\bar{p}$ invariant-mass range from 3.0 to 6.5 GeV/$c^2$ for
events with an undetected ISR photon emitted
close to the collision axis.
From the measured $p\bar{p}$ invariant-mass spectrum we extract the
$e^+e^-\to p\bar{p}$ cross section, and determine the magnitude of the
magnetic form factor of the proton. This is the first measurement of
the proton form factor at $p\bar{p}$ invariant masses higher
than 4.5 GeV/$c^2$. The observed strong decrease of
the form factor for $M_{p\bar{p}} < 4.5$ GeV/$c^2$ agrees with the asymptotic
dependence $\alpha_s^2(M_{p\bar{p}}^2)/M_{p\bar{p}}^4$ predicted by QCD.
There is some indication of an even faster decrease for
$M_{p\bar{p}} > 4.5$ GeV/$c^2$.  

The branching fractions for the decays $J/\psi\to p\bar{p}$ and
$\psi(2S)\to p\bar{p}$ have been measured, and the values
\begin{eqnarray}
{\cal B}(J/\psi\to p\bar{p})=(2.33\pm0.08\pm0.09)\times 10^{-3},\nonumber \\
{\cal B}(\psi(2S)\to p\bar{p})=(3.14\pm0.28\pm0.18)\times 10^{-4}
\end{eqnarray}
have been obtained. These values are in agreement with previous measurements.

\section{ \boldmath Acknowledgments}
We are grateful for the 
extraordinary contributions of our \pep2\ colleagues in
achieving the excellent luminosity and machine conditions
that have made this work possible.
The success of this project also relies critically on the 
expertise and dedication of the computing organizations that 
support \babar.
The collaborating institutions wish to thank 
SLAC for its support and the kind hospitality extended to them. 
This work is supported by the
US Department of Energy
and National Science Foundation, the
Natural Sciences and Engineering Research Council (Canada),
the Commissariat \`a l'Energie Atomique and
Institut National de Physique Nucl\'eaire et de Physique des Particules
(France), the
Bundesministerium f\"ur Bildung und Forschung and
Deutsche Forschungsgemeinschaft
(Germany), the
Istituto Nazionale di Fisica Nucleare (Italy),
the Foundation for Fundamental Research on Matter (The Netherlands),
the Research Council of Norway, the
Ministry of Education and Science of the Russian Federation, 
Ministerio de Ciencia e Innovaci\'on (Spain), and the
Science and Technology Facilities Council (United Kingdom).
Individuals have received support from 
the Marie-Curie IEF program (European Union) and 
the A. P. Sloan Foundation (USA).

\end{document}